# Context, Credibility, and Control: User Reflections on AI-Assisted Misinformation Tools

Varun Sangwan
varun.sangwan@aalto.fi
Aalto University
Espoo, Finland

Heidi Mäkitalo
heidi.makitalo@aalto.fi
Aalto University
Espoo, Finland

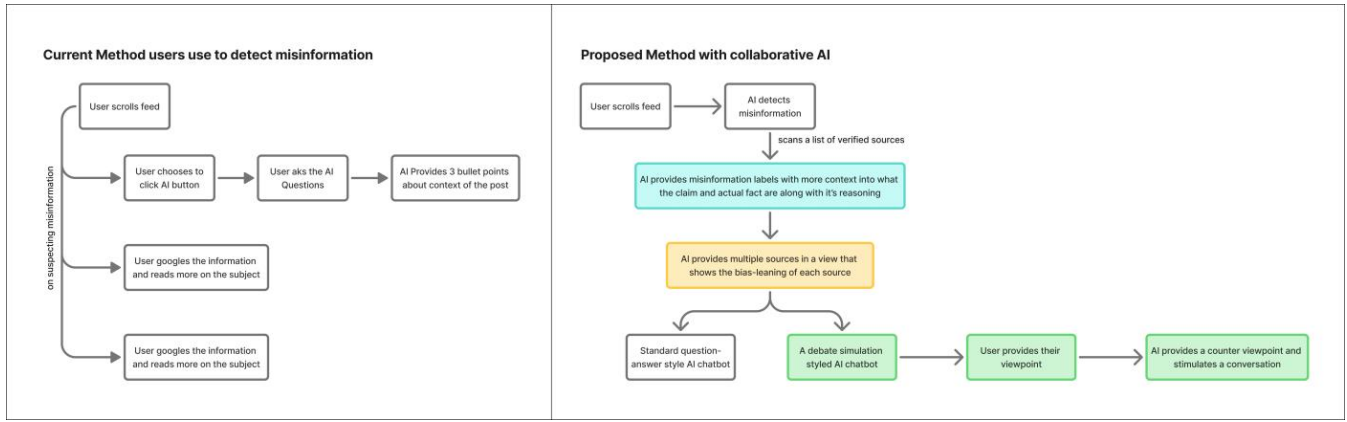

Figure 1: A Comparative Overview of Current and Proposed Methods for Misinformation Detection

## Abstract

This paper contributes to the understanding of how collaborative AI can enhance user agency in identifying and evaluating misinformation on social media platforms. Users typically rely on personal judgment or surface-level fact-checks to determine content credibility. However, these methods are often insufficient in the face of complex, emotionally charged, or context-lacking information—revealing a gap in both design and knowledge surrounding real-time, assistive verification mechanisms.

In response, we developed and tested an interface integrating collaborative AI features—real-time explanations, source aggregation, and debate-style interaction—to help users critically assess content. This improves information discernment by offering contextual cues and argumentative reasoning in a transparent, interactive format. Our user study involving 14 participants shows that 79% found the debate mode more effective than standard chatbots, and the multiple-source view averaged a usefulness rating of 4.6 out of 5.

These findings suggest that interactive, context-rich AI interventions can improve media literacy and user trust in a post-truth digital environment. We conclude that such tools must be ethically designed and user-centered to empower citizens against the rising tide of misinformation.

## 1 Introduction

In today's world, misinformation has emerged as a critical challenge with far-reaching consequences which can vary from influencing elections to public health crises. For example, in February 2024 as many as 36% of people surveyed worldwide by YouGov reported encountering misinformation regarding politics in news media [7]. The spread of false information undermines informed decision-making and erodes trust in institutions. And now because of social media platforms, users are fed content by algorithms that amplify the confirmation bias i.e that expose users to content aligned with their existing beliefs continuously and not hearing the other side of the story thus creating a "filter bubble". As a Human-Computer Interaction (HCI) problem, addressing misinformation requires an understanding of human behavior, algorithmic design, and the social environments in which these systems operate.

The motivation to tackle this issue lies in the current state of social media where users do not critically engage with the content they view. The increasing amount of misinformation on social media platforms exacerbates this issue. Users are prone to being misinformed which creates the need to design collaborative systems that empower social media users to critically engage with content, fostering more informed and balanced perspectives with the content and news they consume.

The central research problem in this study focuses on the limitations of current AI-based solutions for misinformation detection. While existing systems are adept at recognizing patterns in data, they fail to account for the subjective biases of individual users, often reinforcing rather than mitigating these biases. This creates a huge gap in understanding how user-specific biases influence the spread and acceptance of misinformation, and how technology can intervene to counteract this phenomenon. The research question guiding this work is: *How can collaborative AI systems help mitigate misinformation on social media by accounting for user biases and enabling more informed decision-making?*

Figure 1 illustrates the difference between how users currently identify misinformation on social media versus the proposed method using collaborative AI. The left side shows the current approach, where users independently investigate suspicious content by either using generic AI chatbots or manually searching online. This process often lacks depth, personalization, and verified context. In contrast, the right side outlines the proposed collaborative AI method. Here, the AI proactively flags potential misinformation, references a list of verified sources, and provides a clear misinformation label that explains the discrepancy between the claim and the facts. It also presents multiple sources with visible political or ideological leanings to promote transparency. Users can either engage with a standard AI chatbot or participate in a debate-style interaction, where they present their viewpoint and receive a reasoned counterpoint from the AI, stimulating critical thinking and deeper understanding.

Standard AI chatbot interfaces used for fact-checking often function in a linear, question-response format. These systems, while helpful in surfacing additional content or clarifying facts, frequently resemble search engine outputs more than cognitive tools. As a result, they encourage passive interaction: users input a query, receive an answer, and move on. This interaction lacks epistemic depth—offering no space for users to engage critically or reason through competing viewpoints. Moreover, such systems can feel prescriptive or authoritative, which may alienate users with pre existing biases or reinforce skepticism [11].

In contrast, a debate mode interface presents a dialogic, user-driven alternative rooted in active reasoning. Instead of merely consuming answers, users are prompted to formulate arguments, defend positions, or consider opposing perspectives—simulating a cognitive apprenticeship in critical thinking [2]. Studies in educational psychology and learning sciences have shown that structured argumentation improves knowledge retention, mitigates confirmation bias, and fosters intellectual humility [1] [8]. By framing the AI not as a fact-delivering oracle but as a sparring partner in reasoning, the debate mode interface may encourage reflective engagement with contested claims and reduce over reliance on automation.

The proposed solution also leverages collaborative AI to give suggestions and nudges to the user to understand what is going on. This approach integrates the strengths of artificial intelligence in processing vast amounts of data with the nuanced judgment of human users (in most cases - experts) to filter and flag misleading or false information. The challenge is exacerbated by the opacity of current content verification methods. While platforms have introduced fact-check labels and community-based interventions, these are often insufficient. They may be dismissed as partisan or too vague, or fail to provide the reasoning behind the judgment [14].

Our project re-imagines verification as an interactive, educational experience, where users are not told what to believe but are given the resources to explore multiple perspectives and draw their own conclusions.

One of the core challenges in misinformation detection is the lack of adaptability in AI-driven filtering mechanisms, which often operate as static classifiers without considering individual user biases or cognitive processes. Drawing from research on Human-Robot Collaboration (HRC) [9], where machine learning enables robots to interactively adapt to human behaviors, we propose a collaborative AI approach that dynamically engages users in the misinformation detection process. Instead of providing binary fact-checking labels, our system leverages Generative AI to simulate deliberation, nudging users toward critical thinking through context-aware interventions. Inspired by HRC frameworks, which emphasize real-time adaptation and shared decision-making, this approach ensures that AI remains an assistive, rather than authoritative, entity in shaping user engagement with information. This integration of adaptive collaboration principles from HRC into misinformation detection introduces a novel paradigm where AI serves as an interactive partner, guiding but not dictating user judgment, ultimately fostering greater trust and information literacy.

## 2 Literature Survey

Previous work has explored how human judgment in combination with LLM systems could enhance misinformation detection. Zeng et al. (2024) introduce the concept of integrating large language models (LLMs) and crowd sourcing for misinformation detection, aiming to combine the precision of AI with human insight. The paper's primary objective is to evaluate novel combination strategies—Model First, Worker First, and Meta Vote—to enhance the effectiveness of misinformation detection using LLMs and crowd sourced judgments [12]. The key research questions include understanding the performance of LLMs and crowd sourcing individually, identifying when and how a hybrid approach outperforms, and evaluating the robustness of advanced aggregation techniques.

Zhou et al. (2023) focus on the rising challenge of AI-generated misinformation, which exhibits linguistic sophistication and is often indistinguishable from human-created content. The paper investigates the characteristics of AI-generated misinformation compared to human-created misinformation and evaluates the effectiveness of existing misinformation detection models and journalistic guidelines [14]. The research highlights a knowledge gap in understanding how well pre-existing solutions perform on AI-generated misinformation and an ability gap in adapting models to handle the unique linguistic and cognitive features of AI-created content.

The effects of AI-supported misinformation detection on user behavior have also been studied. Jahanbakhsh et al. (2023) explore the effect of combining human assessments and AI predictions in identifying misinformation on social media. In the study, a personalized AI predicts how a user will assess content on Twitter based on the user's previous assessments [3]. The paper finds that the AI predictions influence users' judgment of the (in)accuracy of social media content. This influence seems to grow larger over time, but is mitigated if the user provides reasoning for their assessment [3].

A paper by Lu et al. (2022) looks at the effects of AI-based credibility indicators on people's perceptions of and engagement with the news. The study finds that the presence of AI-based credibility indicators nudges people into aligning their belief in the correctness of news with the AI model's prediction – regardless of the accuracy of this prediction [6]. Therefore, such predictions affect the users' innate accuracy in detecting misinformation. However, AI-based credibility indicators show limited impacts on influencing people's engagement with either real news or fake news when social influence exists, that is, when their judgement of the news is influenced by other people.

Through a large-scale survey, Zhang et al. (2024) find that trust and distrust in social media can coexist in a user rather than being the opposing ends of a spectrum of trust. Specifically, their results show a distinct group of people with both high trust and high distrust. This finding may indicate that a user has trust in specific aspects of the platform and distrust in others [13]. Such coexistence could arise from e.g. the user trusting other users on the platform, but not the platform provider itself [13]. Additionally, the paper finds that both trust and distrust perceptions can shift depending on the platform and are influenced by demographic factors [13]. Finally, while misinformation interventions can elevate users' misinformation awareness and bolster trust in platforms, they don't necessarily reduce distrust.

A significant knowledge gap lies in understanding how AI systems can adapt to user biases, as noted by Zeng et al. (2024) [12] and Zhou et al. (2023) [14], who emphasize the limitations of current solutions in accounting for the linguistic sophistication and variability of misinformation. On the other hand, an ability gap exists in effectively combining AI predictions with human judgment in a manner that mitigates biases rather than amplifies them, as demonstrated by Jahanbakhsh et al. (2023) [3]. These studies also highlight the need for AI to be more transparent and user-aware to influence engagement without compromising innate accuracy. Our paper aims to address these gaps by developing a collaborative AI framework to account for individual biases transparently, while fostering trust and transparency in misinformation detection systems.

In this paper, we propose a collaborative AI framework for bias-aware misinformation detection. Also, unlike traditional AI systems that automatically classify misinformation, our approach fosters deliberative interaction, guiding users toward critical engagement without enforcing censorship. By leveraging computational modeling, user studies, and real-time AI-human collaboration, we aim to create a personalized, adaptable misinformation detection system that enhances user trust, information literacy, and balanced content consumption.

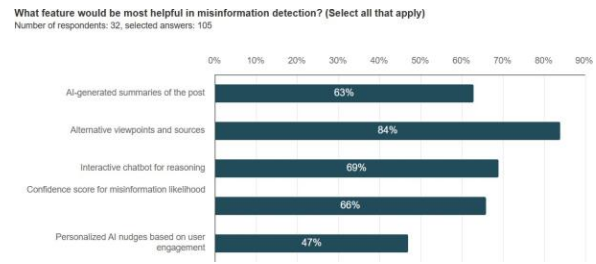

**Figure 2:** *Perceived Helpful Features for Misinformation Detection according to the survey responses*

## 3 Methodology

This research employed a user-centered design methodology supported by usability testing, a strategy well-suited for evaluating experiential features such as interface transparency, explanation clarity, and trust in AI recommendations. While traditional AI benchmarking methods assess backend performance, our focus was on the human experience—how users perceive and interact with AI explanations, and whether these systems can meaningfully support cognitive engagement. A user-focused methodology allows us to capture not just outcomes, but also interpretability, satisfaction, and alignment with users' mental models [10].

### 3.1 User Research

Our pre-study into understanding user behaviour employs a survey-based user research methodology to systematically examine how individuals interact with misinformation and AI-assisted fact-checking tools. Surveys are an effective method for capturing quantitative insights at scale, allowing for the analysis of user perceptions, trust levels, and behavioral tendencies when evaluating online content. This method was chosen due to its ability to collect structured, comparable data across a diverse participant pool while minimizing observational bias.

The survey consisted of close-ended and open-ended questions, focusing on misinformation exposure, fact-checking habits, cognitive biases, and willingness to engage with AI-driven interventions. Participants respond to questions assessing their confidence in identifying misinformation, their tendency to seek alternative viewpoints, and their trust in AI-generated explanations. The goal of this process is to establish a baseline understanding of user behaviors and biases, which was then used to inform the design of a collaborative AI system that enhances critical thinking without enforcing rigid content moderation. The data gathered from this survey provided empirical evidence on how AI interventions should be structured to promote user engagement and cognitive deliberation, ensuring that the system aligns with human decision-making processes rather than overriding them.

The survey results provided critical insights into users' behaviors, attitudes, and expectations regarding AI-assisted misinformation detection. Of the 32 participants that were surveyed, 84.4% of participants reported encountering misinformation on social media either frequently (50%) or almost always (34.4%), yet only 21.9% felt it was "easy" or "very easy" to assess the truthfulness and bias of such content. This underscores a significant gap in users' evaluative skills. When faced with conflicting information, 56.3% indicated they would likely seek alternative viewpoints, showing cognitive openness to understand the viewpoint of other sources of information but also a lack of structured support in that process. Most notably, while 46.9% preferred direct true/false labeling, 40.6% favored AI-assisted reasoning that included contextual explanations and alternative perspectives—highlighting demand for systems that promote active engagement rather than passive consumption.

Support expectations were equally clear: the key features that were rated as most helpful included access to alternative sources (84.4%), interactive reasoning through a chatbot (68.8%), and AI-generated summaries (62.5%) as shown in Figure 2. Furthermore, trust in such tools hinges heavily on transparency (75%) and the inclusion of multiple viewpoints (90.6%), along with the ability to ask follow-up questions (75%). These findings suggest that users are open to seek the advice of AI systems that act as collaborative partners—offering context, inviting user input, and helping develop critical media literacy. Trust, however, remains contingent on the system's transparency, source credibility, and visible reasoning pathways, guiding future design towards interactive, explainable, and user-driven verification frameworks.

### 3.2 Redesigning the interaction for users

The interface was intentionally designed to support users in making informed decisions without feeling overwhelmed or judged. At the heart of the system was a visual label that clearly identified potential misinformation alongside a concise explanation of what made the content questionable. Rather than presenting fact-checks as absolute, the system aimed to be transparent—distinguishing between the claim, the verified fact, and the reasoning behind the AI's decision as shown in Figure 3.

To reduce the feeling of bias or authority, we included multiple sources below each flagged post, sorted across ideological perspectives (left-leaning, centrist, right-leaning). This approach encouraged users to view the issue through different lenses before forming an opinion. Each source also came with a short summary to help users save time while still engaging with opposing views.

We also introduced a feature called "Debate Mode," where the AI would prompt the user to either support or counter the flagged claim. This turned the experience into a mini debate, helping users think more critically by forming arguments, rather than just reading answers. It reframed the AI as a reasoning partner instead of a fact-giver, fostering deeper reflection.

Throughout the interface, the design focused on clarity, transparency, and respect for the user's autonomy. By providing explanations, offering multiple credible sources, and allowing space for reasoning, we aimed to create a tool that didn't just detect misinformation—but helped people understand why it mattered.

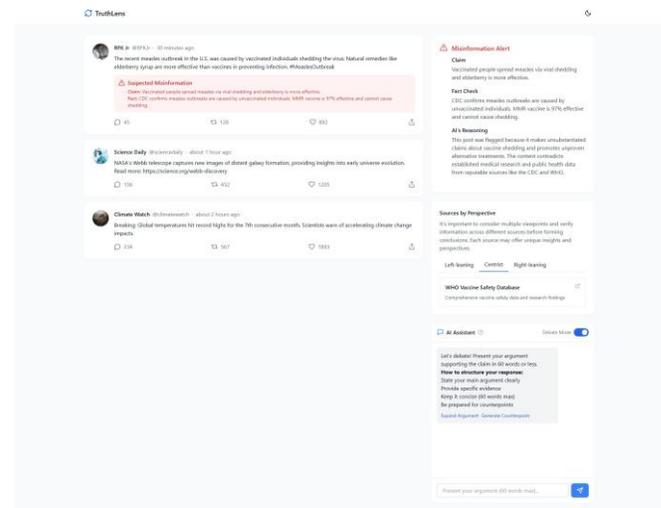

**Figure 3:** *Interface Demonstrating Context-Rich Misinformation Detection and Debate-Based Interaction*

### 3.3 Participants

Fifteen participants were selected from academic and tech-adjacent networks in universities. Selection criteria included regular social media use, fluency in English, and a basic understanding of AI systems. The participants represented a variety of educational and professional backgrounds, ensuring diversity in familiarity with technology and critical thinking approaches. Each participant provided informed consent, and all data was anonymized in accordance with ethical research protocols.

### 3.4 Experiment Design

Participants were invited to complete five sequential tasks on a functional prototype, simulating common social media behaviors. These included reviewing flagged misinformation posts, engaging with the AI's claim-fact breakdown, exploring supporting sources across ideological spectra, chatting with the AI assistant for elaboration, and finally, using the debate mode to construct or challenge arguments. These tasks were selected to reflect realistic user scenarios and test different cognitive demands—from passive recognition to active reasoning. A brief pilot test with two users was conducted to identify UX bottlenecks and calibrate the language of AI outputs.

### 3.5 Pilot Test

A pilot test was conducted using the remote user testing platform Maze to evaluate the effectiveness of the bias-aware collaborative AI prototype. The test involved two participants and focused on how users interact with features like source transparency, debate mode, and bias-tag overlays. Preliminary findings suggested that subtle nudges and intuitive tools are more effective than direct misinformation labels. In terms of how to asses their feedback - users preferred 1–5 scale ratings (likert scale) [4] over binary Yes/No inputs and found remote testing more natural as it allowed them to engage with the system in a way that mirrors typical social media behavior. While the sample size was limited, the test validated the research direction and provided actionable insights for refinement.

## 3.6 Evaluation Metrics

Quantitative metrics included five-point Likert scale ratings for perceived usefulness, clarity, trustworthiness, and novelty. Qualitative metrics included open-ended reflection prompts after each task, enabling users to articulate how and why they did or did not find the tool effective. Metrics were analyzed thematically, using a grounded theory approach to identify recurring sentiments, usability barriers, and feature-specific feedback. This combination of data types enabled a holistic evaluation of both user sentiment and interface efficacy.

## 4 Results

The results indicate a strong user preference for multi-source exploration and debate-style interaction over traditional chatbot-based AI explanations. On average, the users rated usefulness of the provision of additional sources that were sorted according to their bias leaning at 4.6 out of 5. Participants appreciated the categorization of sources by bias and the short summaries provided for each article. This feature was repeatedly cited as "saving time" and offering "an easy way to check things."

The explanation clarity of the standard AI chatbot system similar to X's Grok received an average score of 3.3/5. While users appreciated the claim-fact structure, several participants noted that the language used was too technical or lengthy. Simpler, bite-sized phrasing was requested.

Trust in AI-generated insights remained cautious, with an average trust score of 3.1/5. Some users expressed concern about relying entirely on AI, preferring to conduct personal verification. However, many respondents highlighted that transparency—especially when sources on how the AI got it's information were provided helped to improve it's credibility. The standard AI chatbot (as currently implemented with Grok) interface scored 3.3/5 in usefulness, with criticism centering around redundancy and verbosity and how the content was structured in general.

In contrast to this, the debate mode form of AI Chatbot received a significantly higher rating of 4.0/5 in helping the users gain more information to make a decision. Nearly 79% of users preferred this mode as shown in Figure 4 over the traditional question-answer chatbot for reasoning to identify misinformation, finding it "engaging," "fun," and "more interactive." Several participants noted that arguing for or against a topic activated deeper thought, even if the AI occasionally hallucinated facts. One participant remark sum it up extremely well - *"It was a unique way of presenting an argument. Instead of me asking the question it was refreshing to talk about what I knew about the topic and then use that as the base for the conversation"* . The novelty and dialogic nature of this mode stood out as a preferred method of interaction, particularly for those seeking more than just surface-level information.

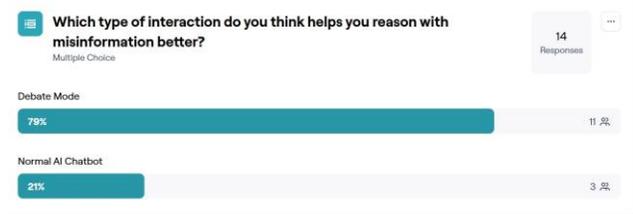

**Figure 4:** *User Preference for Interaction Type in Reasoning with Misinformation*

Qualitative feedback corroborated these insights. Users valued features that helped them understand why something was misinformation rather than merely stating it. Suggestions for improvement included clearer visual cues, shorter chatbot responses, and voice-enabled interfaces. Overall, the results affirm that collaborative, explanatory, and dialogic AI systems can foster more informed and reflective media consumption behaviors.

## 5 Discussion

Our findings indicate that users are not only open to AI assistance in identifying misinformation but actively seek tools that allow for more nuanced, transparent, and participatory experiences. The traditional method of detecting misinformation—relying on users to independently cross-check claims or pose questions to a chatbot—places a high cognitive burden on individuals. In contrast, our proposed method integrates collaborative AI to provide users with richer context, explainable fact-checks, and structured engagement through debate-style dialogue. This approach aligns with recent shifts in HCI that emphasize augmenting human judgment rather than replacing it with automated certainty.

A key insight from our study is the power of debate mode in deepening critical thinking. Participants expressed that being prompted to articulate a viewpoint and immediately encountering a counter-argument helped them better question their assumptions. This aligns with prior research showing that argumentation fosters meta cognition and belief calibration [8]. Rather than passively consuming AI-provided facts, users actively participate in constructing their understanding of what is true or misleading. Notably, this format not only aids in misinformation detection but also cultivates awareness of one's own cognitive biases, such as confirmation bias—a frequent issue in social media environments.

Users appreciated being shown evidence from multiple ideological lenses (left, center, right), which helped them move beyond single-source evaluations and consider how the same claim can be interpreted differently. This design choice echoes calls from journalism and communication research to promote epistemic diversity and resist the effects of echo chambers [5].

In summary, our work highlights that misinformation interventions are most effective when they empower the user—not by spoon-feeding them "truth," but by equipping them with the tools to explore, challenge, and build their own informed perspectives. As AI becomes more embedded in digital communication, designing for collaboration and critical reflection will be key to building media literate societies.

## 6 Conclusion

This study explored how collaborative AI interfaces can support users in identifying and reasoning about misinformation. The results show promise for tools that combine structured context, multiple sources, and dialogic interaction like a debate simulation. While more rigorous testing is needed, our findings support the hypothesis that human-AI collaboration—if well-designed—can reshape how users engage with truth. Future systems must prioritize explainability, interactivity, and user trust to tackle the misinformation crisis at its cognitive root.